\documentclass[aps,twocolumn,showpacs]{revtex4}

\usepackage{dcolumn}
\usepackage{graphicx}

\usepackage{color}

\usepackage{eucal}
\usepackage{mathrsfs}
\usepackage{amsmath}
\usepackage{amssymb}
\usepackage{amsfonts}
\usepackage{latexsym}

\usepackage[hyperindex]{hyperref}

\newcommand{\be}{\begin{eqnarray}}
\newcommand{\ee}{\end{eqnarray}}
\newcommand{\bse}{\begin{subequations}}
\newcommand{\ese}{\end{subequations}}

\newcommand{\bit}{\begin{itemize}}
\newcommand{\eit}{\end{itemize}}
\newcommand{\ben}{\begin{enumerate}}
\newcommand{\een}{\end{enumerate}}

\newcommand{\bpm}{\begin{pmatrix}}
\newcommand{\epm}{\end{pmatrix}}

\newcommand{\mbb}{\mathbb}
\newcommand{\mcal}{\mathcal}
\newcommand{\mfr}{\mathfrak}
\newcommand{\mrm}{\mathrm}

\newcommand{\bs}{\boldsymbol}

\newcommand{\p}{\partial}
\newcommand{\f}{\frac}
\newcommand{\diff}{\mrm{d}}
\newcommand{\lan}{\left\langle}
\newcommand{\ran}{\right\rangle}

\def\dbar{\mathchar'26\mkern-11mu \diff}

\newcommand{\R}{\mbb{R}}

\newcommand{\Z}{\mcal{Z}}

\newcommand{\J}{\mrm{J}}

\newcommand{\ga}{\alpha}
\newcommand{\gb}{\beta}
\newcommand{\gc}{\gamma}

\newcommand{\Gl}{\Lambda}

\newcommand{\veps}{\varepsilon}

\newcommand{\vphi}{\varphi}
\newcommand{\gr}{\varrho}

\newcommand{\Gs}{\Sigma}
\newcommand{\gs}{\sigma}

\newcommand{\gt}{\theta}

\newcommand{\csp}{\;,\qquad\qquad}

\newcommand{\Obs}{\mcal{O}}
\newcommand{\eve}{\mathscr{E}} 

\newcommand{\hyp}{\mfr{H}}  
\newcommand{\cone}{\mfr{C}} 

\newcommand{\iso}{\mfr{I}}  

\newcommand{\kB}{k_\mrm{B}}

\newcommand{\Energy}{\mcal{U}}

\newcommand{\Ent}{\mcal{S}}
\newcommand{\Press}{\mcal{P}}
\newcommand{\Temp}{\mcal{T}}
\newcommand{\Vol}{V}
\newcommand{\Volume}{\mbb{V}}

\newcommand{\Mom}{\mcal{G}}
\newcommand{\Work}{\mcal{A}}
\newcommand{\Heat}{\mcal{Q}}

\newcommand{\diag}{\mrm{diag}}

\newcommand{\dx}{\diff x}

\newcommand{\EW}[1]{\lan{#1} \ran}

\begin{document}

\title{Nonlocal observables and lightcone-averaging in relativistic thermodynamics}

\author{J\"orn Dunkel}
\email{jorn.dunkel@physics.ox.ac.uk}
\affiliation{Rudolf Peierls Centre for Theoretical Physics, University of Oxford, 1 Keble Road, Oxford OX1 3NP, United Kingdom}

\author{Peter H\"anggi}
\affiliation{Institut f\"ur Physik, Universit\"at Augsburg, Universit\"atsstra{\ss}e 1,
D-86135  Augsburg, Germany}

\author{Stefan Hilbert}
\affiliation{Argelander-Institut f\"ur Astronomie, Universit\"at Bonn, Auf dem H\"ugel 71, D-53121 Bonn, Germany}

\date{\today}

\begin{abstract}
The unification of relativity and thermodynamics has been a subject of considerable debate over the last 100 years. The reasons for this are twofold: (i) Thermodynamic variables are nonlocal quantities and, thus, single out a preferred class of hyperplanes in  spacetime. (ii) There exist different, seemingly equally plausible ways of defining heat and work in relativistic systems. These ambiguities led, for example, to various proposals for the Lorentz transformation law of temperature. Traditional \lq isochronous\rq\space formulations of relativistic thermodynamics are neither theoretically satisfactory nor experimentally feasible.  Here, we demonstrate how these deficiencies can be resolved by defining  thermodynamic quantities with respect to the backward-lightcone  of an observation event. This approach yields novel, testable predictions and allows for a straightforward-extension of thermodynamics to General Relativity. Our  theoretical considerations are illustrated through three-dimensional relativistic many-body simulations.
\end{abstract}

\maketitle


\paragraph*{Introduction.--}
Thermodynamics, in the traditional sense, aims at describing
the state of a macroscopic system by means of a few characteristic parameters~$\{Z_i\}$~\cite{1974Ca,Callen,LaLi5,2005Oett}. Typical candidates for thermodynamic state variables~$\{Z_i\}$ are either conserved (extensive)  quantities, e.g. the particle number $N$ and internal energy $U$, or external control parameters that quantify the breaking of symmetries~\cite{1974Ca}. Examples of the latter include the volume $V$ of a confining vessel, indicating the violation of translational invariance, or external magnetic fields, which may break the spatial isotropy. Each extensive state variable is accompanied by an intensive quantity $z_i=\p \Ent/Z_i$, derived from a suitably defined entropy function(al)~$\Ent(\{Z_i\})$. Representing an abstract mathematical theory of differential forms~\cite{2005Oett}, thermodynamic concepts have been successfully applied to vastly different areas, ranging from microscopic many-particle systems~\cite{Callen,LaLi5,2006Ll}, where $\Ent$ is usually interpreted as an information measure (canonical ensemble) or integrated phase space volume (microcanonical ensemble), to exotic objects such as black holes~\cite{1973Be}, where $\Ent$ is related to the black hole's surface area.  

As a coarse-grained macroscopic theory, thermodynamics is inherently nonlocal in that it only considers certain global, or averaged, properties of a physical system~\cite{Callen,LaLi5}. This is rather unproblematic within non-relativistic Newtonian physics, where statements such as "the total energy of a system at time $t$"  are unambiguously defined for arbitrary observers. By contrast -- owing to the absence of a universal time parameter -- the nonlocal character of thermodynamics  has stirred considerable confusion~\cite{1908Pl,1907Ei,1968VK,1963Ott,1966La,1967La,1971TeWe,1995Ko,2009DuHa} in Einstein's theory of relativity~\cite{1948Pr,1967Ga,1970Yu,2007Am}.

To illustrate  the conceptual difficulties in relativistic thermodynamics, consider a confined gas  described by a particle current density $j^\mu(t,\bs x)$ and an energy-momentum tensor density $\gt^{\mu\nu}(t,\bs x)$. If the gas is stationary in some inertial frame $\Gs$, then $j^\mu$  is conserved, i.e., $\p_\mu j^\mu \equiv 0$,  but the divergence of $\gt^{\mu\nu}$ does not identically vanish (due to the pressure arising from the spatial confinement, see example below):
\be\label{e:non}
\p_\mu\gt^{\mu i}\not{\!\!\equiv}\; 0,\qquad
i=1,2,3.
\ee 
This means that space-like surface integrals over $j^\mu$ are independent of the underlying three-dimensional hypersurface $\hyp$ in $(1+3)$-dimensional Minkowski spacetime $\mbb M_4$, whereas those over $\gt^{\mu\nu}$ do depend on $\hyp$. The latter fact is problematic since thermodynamic state variables like energy $\Energy^0$ or momentum $\bs \Energy=(\Energy^1,\Energy^2,\Energy^3)$ are usually defined as surface integrals over the energy-momentum tensor (see App.~\ref{app:notation} and ~\ref{app:integral}  for notation) \cite{1970Yu}: 
\be\label{e:surface_integral}
\Energy^{\nu}[\hyp]&:=&
\int_\hyp \diff \gs_\mu\; \gt^{\mu\nu},\qquad \mu,\nu=0,1,2,3.
\ee
Hence, the first task in relativistic thermodynamics is to identify those hypersurfaces $\{\hyp\}$ that are suitable for defining state variables. Subsequently, one still needs to settle for appropriate definitions of entropy, heat, etc. 

We shall begin by reviewing how these problems are tackled in the most popular, competing versions of relativistic thermodynamics, originally  proposed by Planck~\cite{1908Pl} and Einstein~\cite{1907Ei}, Ott~\cite{1963Ott}, and Van Kampen~\cite{1968VK}, respectively.  A careful analysis elucidates that the traditional approaches are neither conceptually satisfactory nor experimentally feasible. The deficiencies can be cured by defining thermodynamic quantities in terms of lightcone averages. To clarify these aspects, we consider a weakly interacting relativistic gas~\cite{1911Ju}. Notwithstanding, the main conclusions apply to any confined system that can be described by tensor densities $j^\mu,\gt^{\ga\gb},\ldots$. In the second part, we shall discuss observable consequences such as the apparent drift of distant objects that are, in fact, at rest relative to the observer. This surprising effect -- which should be accounted for when estimating the velocities of very hot astrophysical objects from photographic data -- will be illustrated by relativistic many-particle simulations.

\paragraph*{Model (J\"uttner gas).--} 
We consider an enclosed, dilute gas consisting of $N$ relativistic particles [rest mass $m$; velocity $\bs v$; momentum $\bs p=m\bs v(1-\bs v^2)^{-1/2}$; speed of light $c=1$]. Let us assume the gas is  stationary in the \mbox{(\lq lab\rq-)}\-frame $\Gs$,  and can be described by a $\Gs$-time-independent, normalized one-particle phase space probability density function (PDF) 
\bse\label{e:juttner}
\be\label{e:juttner-a}
f(t,\bs x,\bs p)=\vphi(\bs x,\bs p)=\gr(\bs x)\;\phi_\J(\bs p),
\ee 
with J\"uttner momentum distribution~\cite{1911Ju,2007CuEtAl} 
\be\label{e:juttner-b}
\phi_\J(\bs p)= \Z^{-1} \exp(-\gb p^0),\qquad \gb>0.
\ee
$\Z=4\pi m^3\; K_2(\gb m)/(\gb m)$ is the normalization constant, $K_n(z)$ the $n$th modified Bessel function of the second kind~\cite{AbSt72}, and $p^0=(m^2+\bs p^2)^{1/2}$ the particle energy. Later on, the distribution parameter $\gb$ will be identified with the inverse thermodynamic (rest) temperature of the gas. The exact functional form of the spatial density $\gr$ in Eq.~\eqref{e:juttner-a} is irrelevant, as long as $\gr$ is normalizable  (i.e., restricted to a finite spatial volume set $\Volume\subset \R^3$ in $\Gs$).
For simplicity, we may consider a spatially homogeneous gas enclosed in a stationary cubic box $\Volume=[-L/2,L/2]^3$ in $\Gs$, corresponding to
\be\label{e:rho_boxed_gas}
\gr(\bs  x)= 
\begin{cases}
V^{-1},&\text{ if } \bs x\in\Volume,\\
0,&\text{ if } \bs x\not\in\Volume.
\end{cases}
\ee
\ese
Here, $V=L^3$ is the $\Gs$-simultaneously measured (Lebesgue) volume of $\Volume$ in $\Gs$.

The phase space PDF $f$ is a Lorentz scalar~\cite{1969VK}. Thus the current density $j^\mu$ and energy-momentum tensor $\gt^{\mu\nu}$ can be constructed from $f$ by:
\bse
\be\label{e:4-density-covariant}
j^\mu(t,\bs x)&=&
N\int \f{\diff^3 p}{p^0}\; f\;p^\mu,\\ 
\label{e:stress-tensor-covariant}
\gt^{\mu\nu}(t,\bs x)&=& 
N \int \f{\diff^3 p}{p^0}\; f\;p^\mu p^\nu.
\ee
\ese
Concretely,  for Eqs.~\eqref{e:juttner} we have
$(j^\ga)=(\gr,\bs 0)$ and
\be\label{a-e:stress-tensor-covariant-gas}
\gt^{\mu\nu}
=
N\,\gr
\begin{cases} 
\EW{p^0},    &\qquad\mu=\nu=0,\\
\gb^{-1},    &\qquad\mu=\nu=1,2, 3,\\
0,& \qquad\mu\ne\nu,
\end{cases}
\ee
where $\EW{p^0}={3}{\gb^{-1}}+m\;{K_1(\gb m)}/{K_2(\gb m)}$ is the mean energy per particle.
One readily verifies that $\p_\ga j^\ga\equiv 0$, whereas $\p_\mu\gt^{\mu i}=\gb ^{-1}\p_i\gr\ne 0$ at the boundary of $\Volume$, in agreement with Eq.~\eqref{e:non}. Confinement generates stress -- the importance of this seemingly trivial statement shall be seen immediately.

\paragraph*{Isochronous state variables.--}
\begin{figure}[t]
\centering
\includegraphics[width=\linewidth]{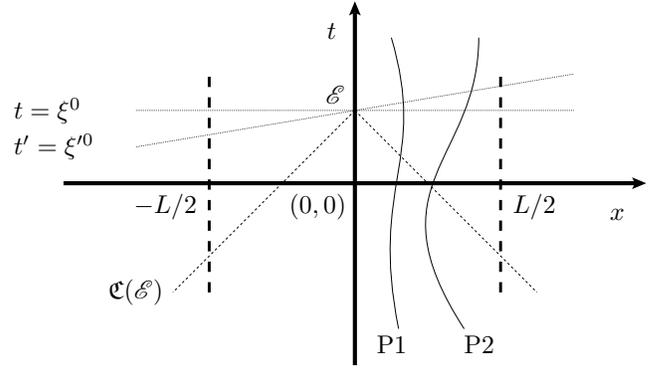}
\caption{
\label{fig01}
The different hypersurfaces as defined in Eq.~\eqref{e:isochronous} and~\eqref{e:lightcone}. The worldlines of two particles are labeled by ``P1'' and ``P2''; the worldlines of the container walls correspond to vertical lines at $x=-L/2$  and $x=L/2$ (dashed lines), respectively.
Assume a lab observer, resting at position $x=0$ in $\Gs$, takes a photograph of the system at the spacetime event $\eve$ with coordinates $(t,x)=(\xi^0,0)$ in~$\Gs$. This photograph will reflect the state of the system along the backward-lightcone $\cone(\eve)$.
}
\end{figure}
The traditional versions of relativistic thermodynamics~\cite{1908Pl,1907Ei,1968VK,1963Ott,1966La,1967La,1995Ko} are recovered from  Eqs.~\eqref{e:juttner}--\eqref{a-e:stress-tensor-covariant-gas} by inserting isochronous spacetime hypersurfaces into Eq.~\eqref{e:surface_integral}. To see this, consider an inertial frame  $\Gs'$, moving at velocity $w$ along the $x^1$-axis of the lab-frame $\Gs$. An event $\eve$ with coordinates $(\xi^0,\bs \xi)$ in $\Gs$ and  $(\xi'^0,\bs \xi')$ in $\Gs'$ defines isochronous hyperplanes $\iso(\xi^0)$ and  $\iso'(\xi'^0)$  in $\Gs$ and $\Gs'$, respectively, by
\bse\label{e:isochronous}
\be
\iso(\xi^0)&:=&\{\;(t,\bs x)\;|\; t = \xi^0\; \},\\
\iso'(\xi'^0)&:=&\{\;(t',\bs x')\;|\; t' = \xi'^0\; \}.
\ee
\ese
If $\Gs$ and $\Gs'$ are in relative motion, these hyperplanes differ from each other, $\iso(\xi^0)\ne\iso'(\xi'^0)$, see Fig.~\ref{fig01}. Inserting $\hyp=\iso[\xi^0]$ into Eq.~\eqref{e:surface_integral}, we obtain the \emph{lab-isochronous} energy-momentum vector $\Energy^\mu[\iso]$ in $\Gs$:
\be\label{e:em-lab}
(\Energy^\mu[\iso])
=
N
(\EW{p^0},\bs 0).
\ee
On the other hand, choosing $\hyp=\iso'[\xi'^0]$ yields the $\Gs'$-isochronous energy-momentum vector  $\Energy'^\mu[\iso']$ in $\Gs'$: 
\bse\label{e:em-moving-moving}
\be\label{e:em-moving}
\Energy'^\mu[\iso']
=
N
\begin{cases}
\gc (\EW{p^0}+w^2 \gb^{-1}),&\qquad \mu=0,\\
-\gc w(\EW{p^0}+ \gb^{-1} ),&\qquad \mu= 1,\\
0,                   &\qquad \mu>1,
\end{cases}
\ee
where $\gc:=(1-w^2)^{-1/2}$. Applying a Lorentz transformation with $-w$ to Eq.~\eqref{e:em-moving}, we find the corresponding energy-momentum in $\Gs$
\be\label{e:em-moving-b}
\Energy^\mu[\iso']
= 
N(\EW{p^0}, -w \gb^{-1}, 0,0).
\ee
Hence, the energy-momentum vectors~\eqref{e:em-lab} and \eqref{e:em-moving} are not related by a Lorentz transformation. In fact,  $\Energy^\mu[\iso]$ and $\Energy^\mu[\iso']$ are connected by
\be
(\Energy^\mu[\iso'])=(\Energy^\mu[\iso]) + N\beta^{-1}(0, w, 0, 0),
\ee
\ese
reflecting the underlying hypersurface and observer velocity. As mentioned earlier, the difference between $\Energy'^\mu[\iso']$ and $\Energy^\mu[\iso']$ arises because the energy-momentum tensor of a spatially confined gas is \emph{not} conserved. It is also a reason for the existence of various temperature Lorentz transformation laws.

\paragraph*{Entropy.--}
Having at hand the state variables \lq energy\rq~$\Energy^0$ and \lq momentum\rq~$\bs\Energy$, one still needs \lq entropy\rq. For a J\"uttner gas, one can define the entropy density four-current~\cite{2008De,CercignaniKremer} by  (units $\kB=1$)
\be\label{e:entropy-4-current}
s^\mu(t,\bs x)=
-N \int \f{\diff^3p}{p^0}\;p^\mu\; 
f \ln(h^3 f).
\ee
The specific shape~\eqref{e:entropy-4-current} is tightly linked to the exponential form of the J\"uttner distribution~\eqref{e:juttner}. In fact, this combination ($f,s^\mu$) is  just one amongst several probabilistic models of thermodynamics; i.e.,  there exist other pairings, e.g., 
based on Renyi-type entropies, that yield consistent thermodynamic relations as well~\cite{2007Ca}.  However,  inserting \eqref{e:juttner} into~\eqref{e:entropy-4-current}, we find
\be\label{e:entropy-4-current-juttner}
s^\mu(t,\bs x)= 
N\gr
\begin{cases}
\ln(\Vol \Z /h^{3})+\gb\EW{p^0}, &\quad\nu=0,\\
0,  &\quad\nu>0.
\end{cases}
\ee
Hence, the current~\eqref{e:entropy-4-current-juttner} is stationary in $\Gs$ and satisfies the conservation law 
\be\label{e:entropy_conservation}
\p_\nu s^\nu\equiv 0.
\ee
The associated thermodynamic entropy $\Ent$ is obtained by integrating $s^\mu$ over some space-like or light-like hyperplane $\hyp$, yielding the Lorentz invariant quantity
\be\label{e:entropy_definition}
\Ent[\hyp] 
:= 
\int_{\hyp} \diff \gs_\nu\; s^\nu(t,\bs x).
\ee
Equation~\eqref{e:entropy_conservation} implies that the integral~\eqref{e:entropy_definition} is the same for the hyperplanes $\iso(\xi^0)$ and $\iso'(\xi'^0)$, 
\be
\Ent[\iso]=\Ent'[\iso]=
\Ent[\iso']=\Ent'[\iso'].
\ee
Thus, there is little or no room for controversy about the transformation laws of entropy in this example. 
The integral~\eqref{e:entropy_definition} is most conveniently calculated along $\hyp=\iso[\xi^0]$ in $\Gs$, yielding
\be\label{e:juttner_S}
\Ent
=
\int\diff^3x\; s^0 
=
N\ln(\Vol \Z/ h^{3} )+\gb N\EW{p^0}.
\ee
This can also be rewritten as
\be
\Ent'&=&\label{e:juttner_entropy}
N \ln (\gc V'\Z/\hbar^3) +\gb\Energy'^0[\iso]/\gc\\
&=&\notag
N \ln (\gc V'\Z/\hbar^3) +\gb\gc(\Energy'^0[\iso'] +w\Energy'^1[\iso'] ),
\ee
where $V'=V/\gc$ denotes the Lorentz contracted (i.e., $\Gs'$-simultaneously measured) volume~\footnote{More precisely, one should write $V'=V'[\iso']$ and $V=V[\iso]$ in order to reflect how volume is measured (defined) in either frame.}.

\paragraph*{Einstein-Planck theory.--}
We are now ready to summarize the most common versions of relativistic thermodynamics. Planck~\cite{1908Pl} and Einstein~\cite{1907Ei} propose to use the $\Sigma'$-synchronous four-vector  $\Energy'^\mu[\iso']$ from Eq.~\eqref{e:em-moving} as thermodynamic energy-momentum state variables. Furthermore, they choose to \emph{define}  heat $\Heat'[\iso']$ and, thus, temperature $\Temp'$ in $\Gs'$ by the following \emph{postulated} form of first law of  thermodynamics~\footnote{Cf.\ Eq.~(23) in \cite{1907Ei}.} 
\bse\label{e:planck}
\be\label{e:planck-a}
\dbar \Heat'[\iso']:=\Temp'\diff\Ent':=
\diff \Energy'^0[\iso']  -w'\diff \Energy'^1[\iso']+ \Press'\diff\Vol',
\quad
\ee
where the intensive variable $w'=-w$ is the constant $x'^1$-velocity of the gas (container) in  $\Sigma'$, and $\Press'$ the pressure. Considering  the special case $w'=0$ first, we see that Eq.~\eqref{e:planck-a} is consistent with the second line of Eq.~\eqref{e:juttner_entropy} upon identifying $\Temp=\gb^{-1}$ and $\Press\Vol=N\gb^{-1}$; i.e., the parameter $\gb$ of the J\"uttner distribution equals the inverse rest  temperature.  Furthermore, for moving systems with $w'\ne 0$, we find that thermodynamic quantities in $\Gs$ and $\Gs'$ are related by \footnote{The scalar transformation law of the pressure is consistent with the Lorentz transformation laws of force and area~\cite{1968VK}.}
\be
\label{e:planck-b}
\Vol'&=&\Vol/\gc,\qquad
\Press'=\Press,\qquad
\Ent'=\Ent,
\qquad\\
\label{e:planck-c}
\Energy'^0[\iso']&=&\gc\left(\Energy^0[\iso]+w'^2\,\Press \Vol\right),\\
\Energy'^1[\iso']&=&\gc w'\left(\Energy^0[\iso]+\Press \Vol\right),\\
\label{e:planck-d}
\Temp'&=&\gc^{-1}\,\Temp=(1-w'^2)^{1/2}\,\Temp,
\ee
\ese
i.e., within the Einstein-Planck formalism  \emph{a moving body appears cooler} \footnote{It seems that, in the later stages of his life, Einstein changed his opinion about the correct transformation laws of thermodynamic quantities, favoring formulas which were later independently derived by Ott~\cite{1963Ott} and Arzelies~\cite{1965Ar,1965Ar_1};  see Liu~\cite{1992Liu,1994Liu}}. Equations~\eqref{e:planck} were criticized in a posthumously published paper by Ott~\cite{1963Ott} and, later, also by van Kampen~\cite{1968VK,1969VK_2} and Landsberg~\cite{1966La,1967La}.

\paragraph*{Ott's and van Kampen's theory.--} 
Ott~\cite{1963Ott} and Van Kampen~\cite{1968VK}  choose to formulate thermodynamic relations in the moving frame  $\Gs'$ in terms of the $\Gs$-isochronous energy-momentum vector $\Energy'^{\mu}[\iso]={\Gl^\mu}_\nu \Energy^{\nu}[\iso]$. They differ, however,  as to how heat and work should be defined.   Van Kampen~\cite{1968VK,1969VK_2} replaces Planck's version of the first law, Eq.~\eqref{e:planck-a}, by introducing a covariant thermal energy-momentum transfer four-vector $\Heat^\mu$ via
\be
\label{e:vk}
\label{e:vk-a}
\dbar\Heat^\mu[\iso]:=\diff \Energy^\mu[\iso]-\dbar\Work^\mu[\iso],
\ee
where, in the (lab) frame $\Gs$, the non-thermal work vector $\Work^\mu[\iso]$ is determined by $(\dbar\Work^\mu[\iso]):=(-\Press \diff \Vol,\bs 0)$.  Accordingly, in a moving frame $\Gs'$ one then finds
$\dbar\Heat'^\mu[\iso]=\diff \Energy'^\mu[\iso]-\dbar\Work'^\mu[\iso]$, where by means of a Lorentz transformation
\be\label{e:vk-e}
\diff \Energy'^\mu[\iso]
=
w'^\mu\;
\diff\Energy^0[\iso],
\qquad
\dbar\Work'^\mu
=
-w'^\mu\;
\Press \diff \Vol.
\ee
Here, $(w'^\mu)=(\gc,\gc w',0,0)$ denotes the velocity four-vector of the gas (container) in~$\Gs'$.  While essentially  agreeing on Eqs.~\eqref{e:vk-a}, \eqref{e:vk-e}, and on the scalar character of entropy $\Ent'=\Ent$, van Kampen and Ott postulate different formulations of the second law, respectively. Specifically, Ott~\cite{1963Ott}  defines the temperature $\Temp'$ in $\Gs'$ via
\bse\label{e:ott-temperature}
\be\label{e:ott-a}
\Temp'\,\diff\Ent'
:=\dbar \Heat'^0
=\gc\, \dbar \Heat^0
= \gc\;\Temp\,\diff\Ent,
\ee
which implies the modified temperature transformation law \footnote{See also Eddington~\cite{1923Ed} and Arzelies~\cite{1965Ar}.}
\be\label{e:ott-b}
\Temp'=\gc\,\Temp=(1-w'^2)^{-1/2}\,\Temp;
\ee
\ese
i.e., according to Ott's definition of heat and temperature \emph{a moving body appears hotter}. 
Van Kampen~\cite{1968VK} argues that the Eqs.~\eqref{e:ott-temperature} are not well-suited if one wishes to describe heat and energy-momentum exchange between systems that move at different velocities (hetero-tachic processes). To achieve a more convenient description, he proposes to characterize the heat transfer by means of a heat scalar $\Heat'=\Heat$, defined by~\cite{1968VK,1969VK_2} 
\bse\label{e:vk-temperature}
\be
\dbar \Heat'
:=-w'_\mu \dbar\Heat'^\mu
=-w_\mu \dbar\Heat^\mu
=\dbar \Heat=\dbar\Heat^0.
\ee
He then goes on to define temperature with respect to $\Heat$,
\be
\Temp' \diff \Ent':=\dbar \Heat'=\dbar \Heat=\Temp \diff \Ent,
\ee
yielding yet another temperature transformation law:
\be
\Temp'=\Temp;
\ee
\ese
i.e., according to van Kampen's definition \emph{a moving body appears neither hotter nor colder}. Adopting this convention, one can define an inverse temperature four-vector $\gb'_\mu:=\Temp'^{-1}\,w'_\mu=\Temp^{-1}\,w'_\mu$ and rewrite the second law in the compact covariant form
\be
\diff\Ent' 
=
-\gb'_\mu \dbar\Heat'^\mu.
\ee
\paragraph*{Discussion.--}
Evidently,  whether a moving body appears hotter or not depends solely on how one defines thermodynamic quantities. The formalisms of Ott~\cite{1963Ott} and van Kampen~\cite{1968VK,1969VK_2} are based on the same \mbox{(lab-)}\linebreak 
isochronous
hyperplane $\iso$ -- they merely differ in their respective temperature definitions~\cite{1970Yu}. By contrast, the Einstein-Planck theory~\cite{1907Ei,1908Pl} is based on an observer-dependent isochronous hyperplane $\iso'$. While, in principle, there is nothing wrong with this, a conceptual downside of the latter approach is that the state variables energy and momentum, when measured in different frames, are not connected by  Lorentz transformations -- or, put differently: To experimentally determine, e.g., the energies $\Energy^0[\iso]$ and $\Energy'^0[\iso']$, two observers  need to perform non-equivalent measurements~\cite{1967Ga}, since measurements must be performed $\Gs$-simultaneously in the first case, but   $\Gs'$- simultaneously in the second case. This might seem sufficient for regarding either Ott's or van Kampen's (more elegant) approach as preferable. However, before adopting this point of view,  it is worthwhile to ask:  
\begin{itemize}
\item \emph{Which formulation is feasible from an experimental point of view?} 
\item \emph{Which formalism provides a suitable conceptual framework for extensions to general relativity~\cite{MiThWe00,Weinberg}?}
\end{itemize}
Unfortunately, from an objective perspective, neither of the above proposals fulfils these criteria. The reason is that either formulation is based on simultaneously defined averages. On the one hand, this means that it is virtually impossible to  
directly measure the quantities appearing in the theory; e.g., in order to determine $\Energy^0[\iso]$ one would have to determine the velocities of the particles at time $t=\xi^0$ in $\Gs$, which requires either superluminal information transport~\cite{1967Ga} or unrealistic efforts of trying to reconstruct isochronous velocity data from recorded trajectories. On the other hand, it is very difficult, if not impossible, to transfer the concept of global isochronicity to general relativity due to the absence of global inertial frames in curved spacetime.

\paragraph*{\lq Photographic\rq~ thermodynamics.--}
To overcome these drawbacks, we propose to define relativistic thermodynamic quantities by means of surface integrals along the backward-lightcone $\cone[\eve]$, where $\eve$ is the event of the observation, see Fig.~\ref{fig01}. This is motivated by the following facts: (i)~A photograph taken by an observer $\Obs$ at the event $\eve$  reflects the state of the system along the lightcone $\cone[\eve]$. (ii)~The hyperplane $\cone[\eve]$ is a relativistically invariant object which is equally accessible for any inertial observer; i.e., if another observer $\Obs'$, who  moves relative to $\Obs$, takes a snapshot at the same event $\eve$, then the resulting picture will reflect the same state of the system -- although the `colors' will be different due to the Doppler effect caused by the observers' relative motion~\cite{Weinberg}.  (iii)~The concept of lightcone-averaging can be easily extended to general relativity.  (iv)~In the nonrelativistic limit $c\to\infty$, the lightcone flattens so that photographic measurements become isochronous in any frame in this limit. Thus, lightcone averages appear to be the best-suited candidates if one wishes to characterize relativistic many-particle systems by means of nonlocally defined, macroscopic variables. 
\par
Mathematically, the backward-lightcones $\cone[\eve]$ in $\Gs$ and $\cone'[\eve]$ in  $\Gs'$ are given by
\bse\label{e:lightcone}
\be 
\cone(\eve)
&:=&
\{\;(t,\bs x)\;|\; t = \xi^0-|\bs x-\bs\xi|\;\},\\
\cone'(\eve)
&:=&
\{\;(t',\bs x')\;|\; t' = \xi'^0-|\bs x'-\bs\xi'|\;\}.
\qquad
\ee
\ese 
Unlike the isochronous hyperplanes $\iso(\xi^0)$ and  $\iso'(\xi'^0)$, the lightcones describe the same set of spacetime events, $\cone(\eve)=\cone'(\eve)$. Fixing  $\hyp=\cone(\eve)$ in Eq.~\eqref{e:surface_integral}, we find (see App.~\ref{app:integral})
\bse\label{e:lightcone-average}
\be
\Energy^{0}[\cone]
&=&\label{e:lightcone-average-a}
N\EW{p^0},\\
\Energy^{i}[\cone]
&=&\label{e:lightcone-average-b}
\frac{N}{\gb}\, \int \diff^3x\;
\f{x^i-\xi^i}{|\bs x-\bs\xi|}\;
\gr(\bs x),
\ee
\ese
Unlike $\bs \Energy[\iso]$ and $\bs \Energy[\iso']$, the vector $\bs \Energy[\cone]$ depends on the space coordinates $\bs \xi$ of the observation event $\eve$. Lightcones are Lorentz-invariant objects, implying that $\Energy^\mu[\cone]$ and $\Energy^{\prime\mu} [\cone']$ are directly linked by a Lorentz transformation, i.e., $\Energy^{\prime\mu}[\cone']=\Lambda^{\mu}_{\nu}\Energy^\nu[\cone]$. Moreover, for a spatially homogeneous J\"uttner gas, it is straightforward to compute the entropy $\Ent[\cone]$ as [cf.~Eq.~\eqref{e:juttner_entropy}]
\be
\Ent[\cone]
&=&\notag
N \ln (V\Z/\hbar^3) +\gb\Energy^0[\cone]\\
&=&\notag
N \ln (\gc V'\Z/\hbar^3) +\gb\gc(\Energy'^0[\cone] -w'\Energy'^1[\cone] )\\
&=&\label{e:juttner_entropy_lc}
\Ent'[\cone],
\ee
where, additionally,  $\Ent[\cone]=\Ent[\iso]$ due to the conservation law~\eqref{e:entropy_conservation}. Thus, depending on which definition of heat we choose, we again end up with either Ott's or van Kampen's temperature transformation law. In our opinion, van Kampen's  approach is more appealing as it defines temperature (similar to rest mass)  as an intrinsic property of the  thermodynamic system, whereas Ott's formalism treats temperature as a dynamic quantity similar to the zero-component of the energy-momentum four-vector. 

\paragraph*{Observable consequences.--}
Since, unlike their isochronous counterparts, the state variables $\Energy^\mu[\cone]$ are experimentally accessible, it is worthwhile to discuss implications for present and future astrophysical observations.
Let us assume that an idealized photograph, taken by an observer $\Obs$ at $\eve$, encodes both the positions and velocities (e.g., from Doppler shifts) of a confined gas. If $\Obs$ is at rest relative to the gas, then the mean values of the energy and momentum sampled from the photographic data will converge to $\Energy^\mu[\cone]$ given by Eq.~\eqref{e:lightcone-average}. Equation~\eqref{e:lightcone-average-a} implies that it does not matter for an observer at rest in~$\Gs$ whether energy values are sampled from a photograph or from $\Gs$-simultaneously collected (i.e., reconstructed) data.

The situation is different, when estimating the mean momentum from photographic data. Equation~\eqref{e:lightcone-average-b} implies that the lightcone-average depends on the observer position $\bs \xi$~\footnote{Averages in the lab frame do not depend on the specific value $\xi^0$ of the time coordinate if the PDF is stationary in this frame.}.
A distinguished `photographic center-of-mass' position $\bs \xi_*$ in $\Gs$ can defined by
\be
\Energy^{i}[\cone]\bigr|_{\bs \xi=\bs \xi_*}=0
\csp
i=1,2,3.
\ee 
For example, if $\gr$ is symmetric with respect to the origin of $\Gs$, then $\bs \xi_*=\bs 0$. This would correspond to a lightcone as drawn in  Fig.~\ref{fig01}. In this (and only this) case,  we find $\Energy'^1[\cone] =w'\Energy'^0[\cone] $ and, thus,  lightcone thermodynamics reduces to the Ott-van Kampen formalism. 

To illustrate how $\Energy^{i}[\cone]$ generally depends on the observer's position, let us consider a gas with density profile~\eqref{e:rho_boxed_gas}. For a stationary observer at a position $\bs \xi$ far away from $\Volume$, we can approximate $|\bs x-\bs\xi|\simeq|\bs\xi|$ in Eq.~\eqref{e:lightcone-average-b}, yielding~\footnote{Equations~\eqref{e:lightcone-average} and~\eqref{e:apparent_drift} are  consistent with Eq.~\eqref{e:em-moving-moving}, as can be seen by letting $w\to1$  in~\eqref{e:em-moving-b} and $\xi^1\gg 0$, $\xi^2=\xi^3=0$ in~\eqref{e:apparent_drift}.}
\be
\label{e:apparent_drift}
\Energy^{i}[\cone]=-\f{\xi^i}{|\bs\xi|} N\Temp.
\ee
Hence, a distant observer $\Obs$ who naively estimates $\Energy^{i}[\cone]$ from photographic data could erroneously conclude that the gas is moving away with a momentum vector proportional to the temperature. Reinstating constants $c$ and $\kB$, this relativistic effect becomes neglible if $mc^2\gg \kB\Temp$, but -- given the rapid  improvement of telescopes and spectrographs~\cite{2006Be} -- it should be taken into consideration when estimating the velocities of astrophysical objects from photographs in the future.  In particular, since Lorentz transformations mix energy and spatial momentum components,  for a moving observer $\Obs'$  both $\Energy'^{0}[\cone]$ and $\bs \Energy'^{i}[\cone]$ will be affected, see Fig.~\ref{fig04}. In principle, similar phenomena arise whenever one is limited to photographic observations of partial components in distant compound systems (e.g., the gas in a galaxy), if the energy-momentum tensor of this partial component is not conserved.  At present, it is an open problem whether or not these effects may even be amplified in curved spacetime geometries.

\paragraph*{Numerical results.--}
The preceding theoretical considerations can be illustrated by (1+3)-dimensional relativistic many-body simulations. Compared with the nonrelativistic case, simulations of relativistic many-particle systems are more difficult because particle collisions cannot be modeled by simple interaction potentials anymore~\cite{1949WhFe,1963CuJoSu}.  Generalizing recently proposed lower-dimensional algorithms~\cite{2007CuEtAl,2008MoGhBa}, our computer experiments are based on hard-sphere-type interactions in the two-particle center-of-mass frame (see App.~\ref{app:simulations} details). This model is fully relativistic in the low-to-intermediate density regime~\footnote{Generally, hard-sphere models become \lq nonrelativistic\rq~at very high densities (liquid/solid regime).}.
\begin{figure}[t]
\centering
\includegraphics[width=\linewidth]{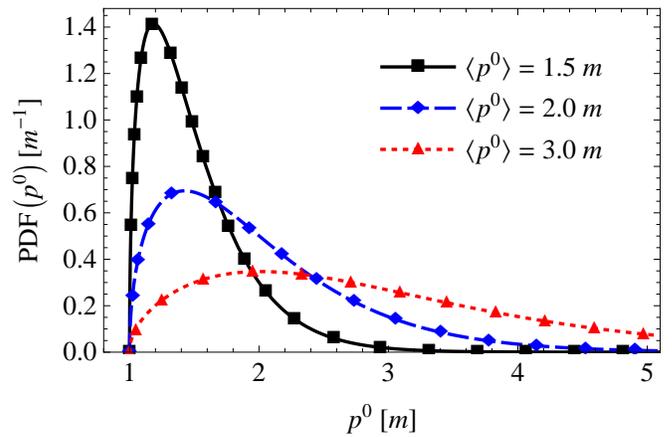}
\caption{
\label{fig:fig02}
The distribution of the particle energy $p^0$ in the rest frame of the gas for various values of the mean particle energy $\EW{p^0}$: The distribution measured in our simulations (symbols) is compared to the J{\"u}ttner distribution~\eqref{e:juttner-b} (lines).
}
\end{figure}
Figures~\ref{fig:fig02}--\ref{fig04} depict results of simulations with $N=1000$ particles.  Initially, all particles are randomly distributed in a cubic box with same energy $p^0$, but random velocity directions. After a few collisions per particle, the energy distribution relaxes to the J{\"u}ttner distribution~\eqref{e:juttner-b}, see Fig.~\ref{fig:fig02}, which confirms that our collision algorithm works correctly in this density regime.
\begin{figure}[t]
\centering
\includegraphics[width=\linewidth]{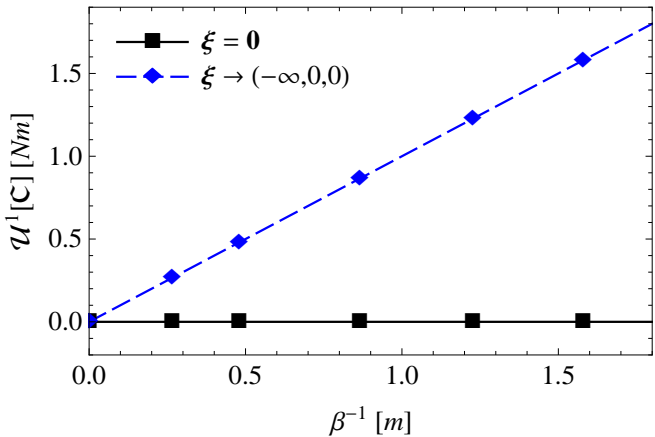}
\caption{
\label{fig:fig03}
The lightcone momentum $\Energy^1[\cone]$ as a function of gas temperature $\beta^{-1}$ for an observer resting either at the center of the gas container, $\bs \xi=\bs 0$, or far away from it, $\bs \xi\to(-\infty,0,0)$. Symbols indicate values measured in the simulations, and lines the theoretical predicted curves.
}
\end{figure}
The thermodynamic energy-momentum vector $\Energy^{\prime\mu}[\hyp]$ is determined by recording each particles' momentum as its trajectory passes through the corresponding hyperplane.  We first consider an observer $\Obs$ who is at rest relative to the gas.  As predicted by Eq.~\eqref{e:apparent_drift} we find that, if the location of $\Obs$ deviates from the center of the box, a photo made by $\Obs$ yields a \emph{non-zero} momentum $\Energy^i\propto \gb^{-1}$, see blue line in Fig.~\ref{fig:fig03}.
For comparison, Fig.~\ref{fig04} shows the results for a moving observer  $\Obs'$ (speed $w$), obtained by isochronous sampling along different hyperplanes $\iso'(w)$, or photographic sampling along the backward-lightcones $\cone(\eve)$. Again, as predicted by the theory, the resulting overall energy-momentum does not only depend on the observer velocity, but also on the underlying hypersurface and, in particular, on the observer's position. Although our study still neglects quantum processes and gravity, which play an important role in real astrophysical systems, the results suggest that one needs to be very careful when reconstructing the velocities of hot, relativistic objects from photographic measurements.

\begin{figure}[t]
\centering
\includegraphics[width=\linewidth]{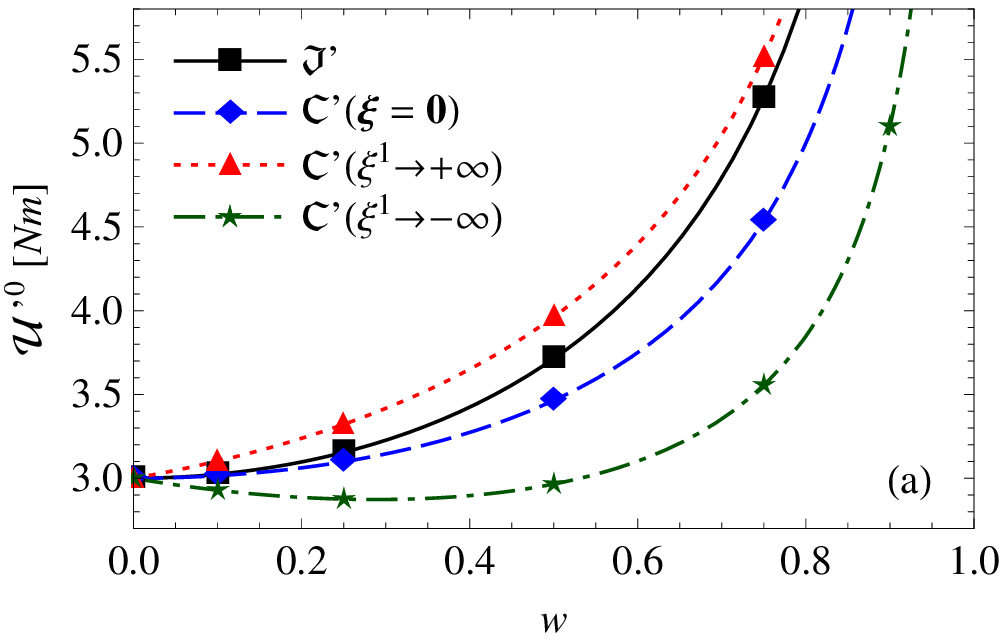}
\includegraphics[width=\linewidth]{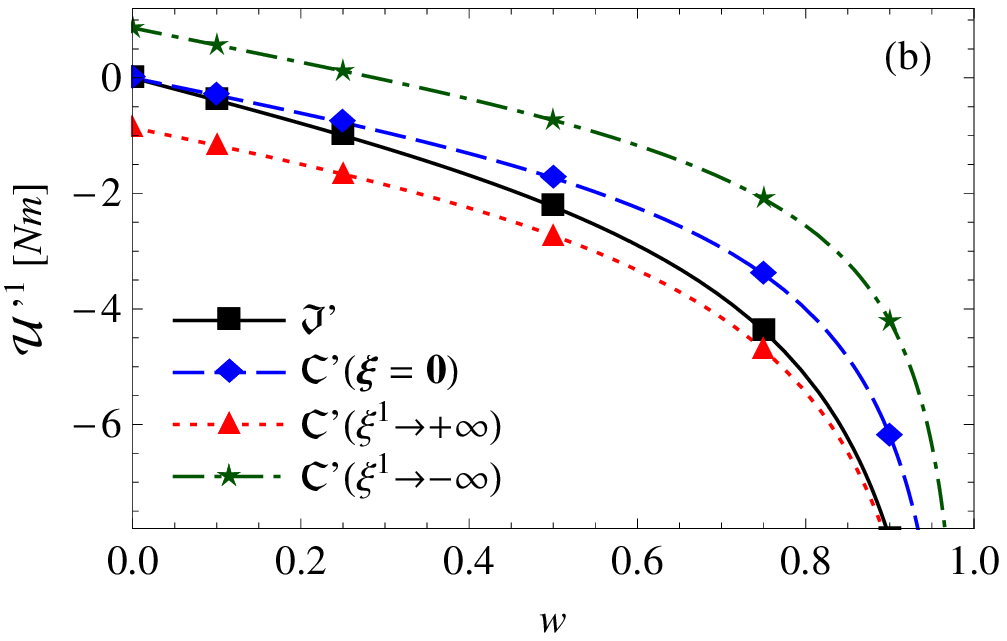}
\caption{
\label{fig04}
Observed energy $\Energy^{\prime0}$ (a) and momentum $\Energy^{\prime1}$ (b) of the gas as a function of the observer speed $w$ along the $x^1$-axis, either sampled from an isochronous hyperplane $\iso'$, or a lightcone $\cone'$. The mean particle energy is $\EW{p^0}=3 m$. The observer using a lightcone is either at the center of the gas container ($\bs \xi=\bs 0$), far behind the container ($\xi^1\to -\infty$), or ahead of the container ($\xi^1\to +\infty$). Simulation results are indicated by symbols, the theoretical prediction by lines. 
}
\end{figure}

\paragraph*{Conclusions.--}
Sometimes, discussions of relativistic thermodynamics start by postulating a set of macroscopic state variables, whose thermodynamics relations (and Lorentz transformations laws)  are subsequently deduced by plausibility considerations. Unfortunately, this approach -- although quite successful in nonrelativistic physics -- is intrinsically limited in a relativistic framework, as it conceals the actual source for conceptual difficulties, namely, the nonlocal character of thermodynamic quantities.  The above analysis may provide guidance for constructing consistent relativistic thermodynamic theories for more complex systems, e.g., based on \mbox{(non-)}\-conserved tensor densities as derivable from classical or quantum Lagrangians.  Care is required when integrating such densities to obtain global thermodynamic state variables, since conservation laws may be violated due to confinement, so that averages may vary depending on the underlying hyperplane(s). Within a conceptually satisfying and experimentally feasible framework, thermostatistical averaging procedures should be defined over lightcones rather than isochronous hypersurfaces.
To put it somewhat provocatively, the isochronous definition of nonlocal quantities, as adopted in traditional formulations of relativistic thermodynamics, can be viewed as a relic of our accustomed nonrelativistic~\lq thinking\rq. With regard to present and future astrophysical observations, it will be important to better understand how the temperature-dependent, apparent drift effects due to lightcone-averaging (i.e., photographic measurements)  become modified in curved spacetime, because this might affect  velocity estimates for astronomical objects which are pivotal for our understanding of the  cosmological evolution~\cite{2006Be}.

\appendix

\section{Notation}
\label{app:notation} 
We adopt units such that speed of light $c=1$ and Boltzmann constant $\kB=1$, and the metric convention $(\eta_{\mu\nu})= \diag(-1,1,1,1)= (\eta'_{\mu\nu})$. Einstein's sum rule is applied throughout. If an event $\eve$ has coordinates $(\xi^\mu)=(\xi^0,\xi^i)=(\xi^0,\xi^1,\xi^2,\xi^3)=(t,\bs \xi )$ in the inertial spacetime frame $\Gs$,   then its coordinates in another frame $\Gs'$, moving at constant relative velocity $w$ along the $x^1$-axis of $\Gs$, are given by $\bigl(\xi'^\mu\bigr)=\bigl(\gamma (\xi^0- w \xi^1),\gamma (\xi^1- w \xi^0), \xi^2, \xi^3\bigr)$ with $\gc=(1-w^2)^{-1/2}$. In short, $\xi'^\mu = {\Gl^\mu}_\nu \xi^\nu$, where $({\Gl^\mu}_\nu)$ is the corresponding Lorentz transformation matrix.

\section{Surface integrals in Minkowski spacetime}
\label{app:integral}
We wish to calculate the integral
\be\label{e:surface_integral_app}
\Mom^{\ga\gb\ldots}[\hyp]&:=&\int_\hyp \diff \gs_\mu\; \gt^{\mu\ga\gb\ldots}(t,\bs x),
\ee
where $\hyp$ is a three-dimensional hyperplane in the $(1+3)$-dimensional Minkowski frame $\Gs$.  If $\gt^{\mu\ga\gb\ldots}$ is a tensor of rank $n$ then $\Mom^{\ga\gb\ldots}[\hyp]$ has rank $(n-1)$. Considering Cartesian spacetime coordinates, the surface element $\diff \gs_\mu$ may be expressed in terms of the alternating differential form~\cite{1967Ha_2}
\be
\label{e:surface_integral_element}
\diff \gs_\mu=
(3!)^{-1}\veps_{\mu\ga\gb\gc}\dx^\ga\wedge \dx^\gb\wedge \dx^\gc,
\ee
where $\veps_{\mu\ga\gb\gc}$ is the Levi-Cevita tensor~\cite{SexlUrbantke}, and $\dx^\ga \wedge \dx^\gb=-\dx^\gb \wedge \dx^\ga$ the antisymmetric product. 
With regard to thermodynamics, we are interested in integrating over space-like or time-like surfaces $\hyp$ given in the form $
x^0=t=g(\bs x).$ Examples are the isochronous hyperplane $\iso(\xi^0)$ from Eq.~\eqref{e:isochronous} and the lightcone $\cone[\eve]$ from  Eq.~\eqref{e:lightcone}. Given the function $g$, we may write $\dx^0=\p_i g\;\dx^i$.
Inserting this expression into Eq.~\eqref{e:surface_integral_app} yields
\be
\Mom^{\ga\ldots}
&=&\notag
\int_\hyp \dx^1\wedge \dx^2\wedge \dx^3\;\left[
\gt^{0\ga\ldots} - 
(\p_i g)\; \gt^{i\ga\ldots} \right]\\
&:=&\notag
\int\diff^3 x
\left[
\gt^{0\ga\ldots} (g(\bs x),\bs x) - (\p_i g)\; \gt^{i\ga\ldots}(g(\bs x),\bs x) \right].
\ee
In particular, for the isochronous hyperplane  $\iso(\xi^0)$  from Eq.~\eqref{e:isochronous}, we have $g(\bs x)=\xi^0$ and $\p_i g=0$ in $\Gs$, leading to
\be
\Mom^{\ga\ldots}[\iso]
&=&
\int\diff^3 x\;\gt^{0\ga\ldots}(\xi^0,\bs x).
\ee
For the lightcone $\cone(\eve)$ with $\p_i g=-(x^i-\xi^i)/|\bs x-\bs\xi|$,  we find
\be\label{a-e:lc-average}
\Mom^{\ga\ldots}[\cone]
&=&\notag
\int \diff^3x\;\biggl\{
\gt^{0\ga\ldots}(\xi^0-|\bs x-\bs\xi|,\bs x) \;+
\\
&&\quad
\f{x^i-\xi^i}{|\bs x-\bs\xi|}\;
\gt^{i\ga\ldots}(\xi^0-
|\bs x-\bs\xi|,\bs x)\biggr\}.
\ee

\section{Numerical simulations}
\label{app:simulations}
Fully relativistic $N$-body simulations would require to also compute the interaction fields generated by particles, which is numerically expensive. For dilute gases with short-range interactions one can obtain reliable results by considering simplified models based on quasi-elastic collisions. In our computer experiments, we simulate a three-dimensional gas of relativistic  hard spheres in a cubic box. In a particle-particle collision, momentum is transferred instantaneously at the moment of closest encounter by taking into account the relativistic energy-momentum conservation laws.

The tasks during a simulation time-step are \cite{2007CuEtAl,2008MoGhBa}: (i) Determine the times/distances of all particle pairs at their closest encounter. (ii) Advance all particles to the next collision time. (iii) Transfer momentum between the colliding particles. (iv) Record particle energies and momenta when the particles are reflected at the walls (e.g., to measure the pressure on the boundaries) or their trajectories pass an observer hypersurface.

Our simulations show that details of the momentum transfer mechanism (e.g., the differential cross-sections) do not affect the stationary momentum distribution. It is, however, important to employ a Lorentz-invariant collision criterion (we use the minimum distance of particles in the two-particle center-of-mass frame). Non-invariant criteria (e.g., the minimum inter-particle distance in the lab frame) may lead to deviations from the J{\"u}ttner distribution.

The most time-consuming task is to determine the closest-encounter times/distances for all particle pairs. A huge speed-up can be achieved by considering only close particle pairs, using a hash table based on a partition of the simulation box into subcubes. With this method one can efficiently simulate $10^3$ particles and $10^6$ collisions in $30\,\text{min}$ on a desktop pc.



\begin{thebibliography}{10}

\bibitem{1974Ca}
H.~Callen.
\newblock Thermodynamics as a science of symmetry.
\newblock {\em Found. Phys.}, 4(4):423--442, 1974.

\bibitem{Callen}
H.~B. Callen.
\newblock {\em Thermodynamics and an Introduction to Thermostatistics}.
\newblock John Wiley \& Sons, New York, 2 edition, 1985.

\bibitem{LaLi5}
L.~D. Landau and E.~M. Lifshitz.
\newblock {\em Statistical Physics}, volume~5 of {\em Course of Theoretical
  Physics}.
\newblock Butterworth-Heinemann, Oxford, 3 edition, 2003.

\bibitem{2005Oett}
H.~C. \"Ottinger.
\newblock {\em Beyond Equilibrium Thermodynamics}.
\newblock Wiley-IEEE, 2005.

\bibitem{2006Ll}
S.~Lloyd.
\newblock Quantum thermodynamics: Excuse our ignorance.
\newblock {\em Nature Phys.}, 2:727--728, 2006.

\bibitem{1973Be}
Jacob~D. Bekenstein.
\newblock Black holes and entropy.
\newblock {\em Phys. Rev. D}, 7(8):2333--2346, Apr 1973.

\bibitem{1908Pl}
M.~Planck.
\newblock {Zur Dynamik bewegter Systeme}.
\newblock {\em Ann. Phys. (Leipzig)}, 26:1--34, 1908.

\bibitem{1907Ei}
A.~Einstein.
\newblock {\"Uber das Relativit\"atsprinzip und die aus demselben gezogenen
  Folgerungen}.
\newblock {\em Jahrbuch der Radioaktivit\"at und Elektronik}, 4:411--462, 1907.

\bibitem{1968VK}
N.~G. {van Kampen}.
\newblock Relativistic thermodynamics of moving systems.
\newblock {\em Phys. Rev.}, 173:295--301, 1968.

\bibitem{1963Ott}
H.~Ott.
\newblock {Lorentz-Transformation der W\"arme und der Temperatur}.
\newblock {\em Z. Phys.}, 175:70--104, 1963.

\bibitem{1966La}
P.~T. Landsberg.
\newblock Does a moving body appear cool?
\newblock {\em Nature}, 212:571--572, 1966.

\bibitem{1967La}
P.~T. Landsberg.
\newblock Does a moving body appear cool?
\newblock {\em Nature}, 214:903--904, 1967.

\bibitem{1971TeWe}
D.~{Ter Haar} and H.~Wergeland.
\newblock Thermodynamics and statistical mechanics on the special theory of
  relativity.
\newblock {\em Phys. Rep.}, 1(2):31--54, 1971.

\bibitem{1995Ko}
A.~Komar.
\newblock {Relativistic Temperature}.
\newblock {\em Gen. Rel. Grav.}, 27(11):1185--1206, 1995.

\bibitem{2009DuHa}
J.~Dunkel and P.~H\"anggi.
\newblock Relativistic {B}rownian motion.
\newblock {\em Phys. Rep.}, 471(1):1--73, 2009.

\bibitem{1970Yu}
C.~K. Yuen.
\newblock Lorentz transformation of thermodynamic quantities.
\newblock {\em Am. J. Phys.}, 38:246--252, 1970.

\bibitem{1948Pr}
M.~H.~L. Pryce.
\newblock The mass-centre in the restricted theory of relativity and its
  connexion with the quantum theory of elementary particles.
\newblock {\em Proc. Roy. Soc. London}, 195(1040):62--81, 1948.

\bibitem{1967Ga}
A.~Gamba.
\newblock Physical quantities in different reference systems according to
  relativity.
\newblock {\em Am. J. Phys.}, 35(2):83--89, 1967.

\bibitem{2007Am}
G.~{Amelino-Camelia}.
\newblock Relativity: Still special.
\newblock {\em Nature}, 450:801--803, 2007.

\bibitem{1911Ju}
F.~J\"uttner.
\newblock {Das {M}axwellsche {G}esetz der {G}eschwindigkeitsverteilung in der
  {R}elativtheorie}.
\newblock {\em Ann. Phys. (Leipzig)}, 34(5):856--882, 1911.

\bibitem{2007CuEtAl}
D.~Cubero, J.~Casado-Pascual, J.~Dunkel, P.~Talkner, and P.~H\"anggi.
\newblock {Thermal equilibrium and statistical thermometers in special
  relativity}.
\newblock {\em Phys. Rev. Lett.}, 99:170601, 2007.

\bibitem{AbSt72}
M.~Abramowitz and I.~A. Stegun, editors.
\newblock {\em Handbook of Mathematical Functions}.
\newblock Dover Publications, Inc., New York, 1972.

\bibitem{1969VK}
N.~G. {van Kampen}.
\newblock Lorentz-invariance of the distribution in phase space.
\newblock {\em Physica}, 43:244--262, 1969.

\bibitem{2008De}
F.~Debbasch.
\newblock Equilibrium distribution function of a relativistic dilute perfect
  gas.
\newblock {\em Physica A}, 387(11):2443--2454, 2007.

\bibitem{CercignaniKremer}
C.~Cercignani and G.~M. Kremer.
\newblock {\em The Relativistic Boltzmann Equation: Theory and Applications},
  volume~22 of {\em Progress in mathematical physics}.
\newblock Birkh\"auser Verlag, Basel, Boston, Berlin, 2002.

\bibitem{2007Ca}
M.~Campisi.
\newblock {Thermodynamics with generalized ensembles: The class of dual
  orthodes}.
\newblock {\em Physica A}, 385(2):501--517, 2007.

\bibitem{1969VK_2}
N.~G. {van Kampen}.
\newblock Relativistic thermodynamics.
\newblock {\em J. Phys. Soc. Jap. Suppl.}, 26:316--321, 1969.

\bibitem{MiThWe00}
C.~W. Misner, K.~S. Thorne, and J.~A. Wheeler.
\newblock {\em Gravitation}.
\newblock W. H. Freeman and Co., New York, 2000.
\newblock 23rd printing.

\bibitem{Weinberg}
S.~Weinberg.
\newblock {\em Gravitation and Cosmology}.
\newblock John Wiley \& Sons, 1972.

\bibitem{2006Be}
C.~L. Bennett.
\newblock Cosmology from start to finish.
\newblock {\em Nature}, 440:1126--1131, 2006.

\bibitem{1949WhFe}
J.~A. Wheeler and R.~P. Feynman.
\newblock {Classical Electrodynamics in Terms of Direct Interparticle Action}.
\newblock {\em Rev. Mod. Phys.}, 21(3):425--433, 1949.

\bibitem{1963CuJoSu}
D.~G. Currie, T.~F. Jordan, and E.~C.~G. Sudarshan.
\newblock {Relativistic Invariance and Hamiltonian Theories of Interacting
  Particles}.
\newblock {\em Rev. Mod. Phys.}, 35(2):350--375, 1963.

\bibitem{2008MoGhBa}
A.~{Montakhab}, M.~{Ghodrat}, and M.~{Barati}.
\newblock {Statistical thermodynamics of a two dimensional relativistic gas}.
\newblock {\em arXiv:0809.1517}, 2008.

\bibitem{1967Ha_2}
R.~Hakim.
\newblock {Remarks on Relativistic Statistical Mechanics. I.}
\newblock {\em J. Math. Phys.}, 8(6):1315--1344, 1965.

\bibitem{SexlUrbantke}
R.~U. Sexl and H.~K. Urbantke.
\newblock {\em Relativity, Groups, Particles}.
\newblock Springer Physics. Springer, Wien, 2001.

\bibitem{1965Ar}
H.~Arzelies.
\newblock {Transformation relativiste de la temperature et de quelques autres
  grandeurs thermodynamiques}.
\newblock {\em Nuovo Cimento}, 35:792--804, 1965.

\bibitem{1965Ar_1}
H.~Arzelies.
\newblock Sur le concept de temperature en thermodynamique relativiste et en
  thermodynamique statistique.
\newblock {\em Nuovo Cimento B}, 40(2):333--344, 1965.

\bibitem{1992Liu}
C.~Liu.
\newblock {Einstein and Relativistic Thermodynamics}.
\newblock {\em Brit. J. Hist. Sci.}, 25:185--206, 1992.

\bibitem{1994Liu}
C.~Liu.
\newblock {Is There A Relativistic Thermodynamics? A Case Study in the Meaning
  of Special Relativity}.
\newblock {\em Stud. Hist. Phil. Sci.}, 25:983--1004, 1994.

\bibitem{1923Ed}
A.~S. Eddington.
\newblock {\em The Mathematical Theory of Relativity}.
\newblock University Press Cambridge, 1923.

\end{thebibliography}
\end{document}